\input{aipcheck}

\documentclass[sort&compress,final]
  {aipproc}

\usepackage{epsfig}

\layoutstyle{6x9}

\def\as{\alpha_s}
\def\asz{\as(\mz)}
\def\mz{M_Z}
\def\etjet{E_T^{\rm jet}}
\def\colab#1{#1 Coll.}
\def\etal{et al.}
\def\Journal#1#2#3#4{{#1} {#2} (#3) #4}

\def\NPB{{\em Nucl. Phys.} {\bf B}}
\def\PLB{{\em Phys. Lett.}  {\bf B}}

\def\PRD{{\em Phys. Rev.} {\bf D}}

\def\EPC{{\em Eur. Phys. Jour.} {\bf C}}

\def\JPG{{\em J. Phys.} {\bf G}}
\def\PS{{\em Phys. Scr.}}

\begin{document}

\title{Precision Measurements of $\as$ at HERA\footnote{Talk given in
    the XIII International Workshop on Deep Inelastic Scattering,
    April $27^{\rm th}$ - May $1^{\rm st}$, 2005, Madison, Wisconsin
    USA.}}

\classification{12.38.Qk}
\keywords{strong coupling constant}

\author{Claudia Glasman\footnote{Ram\'on y Cajal Fellow.}}
{address={Universidad Aut\'onoma de Madrid}
}

\begin{abstract}
The precision measurements of the strong coupling constant, $\as$, and
its energy-scale dependence carried out at HERA by the H1 and ZEUS
Collaborations are reviewed. An average value of \\
\centerline{
$\overline\asz=0.1186\pm 0.0011\ ({\rm exp.})\pm 0.0050\ ({\rm th.})$}\\
is obtained from these measurements. 
The combined HERA determinations of the energy-scale dependence
of $\as$ clearly show the running of $\as$ from jet data alone and are
in agreement with the running of the coupling as predicted by QCD.
\end{abstract}

\maketitle

\section{Introduction}

\begin{figure}[h]
\setlength{\unitlength}{1.0cm}
\begin{picture} (10.0,8.0)
\put (-2.0,0.0){\epsfig{figure=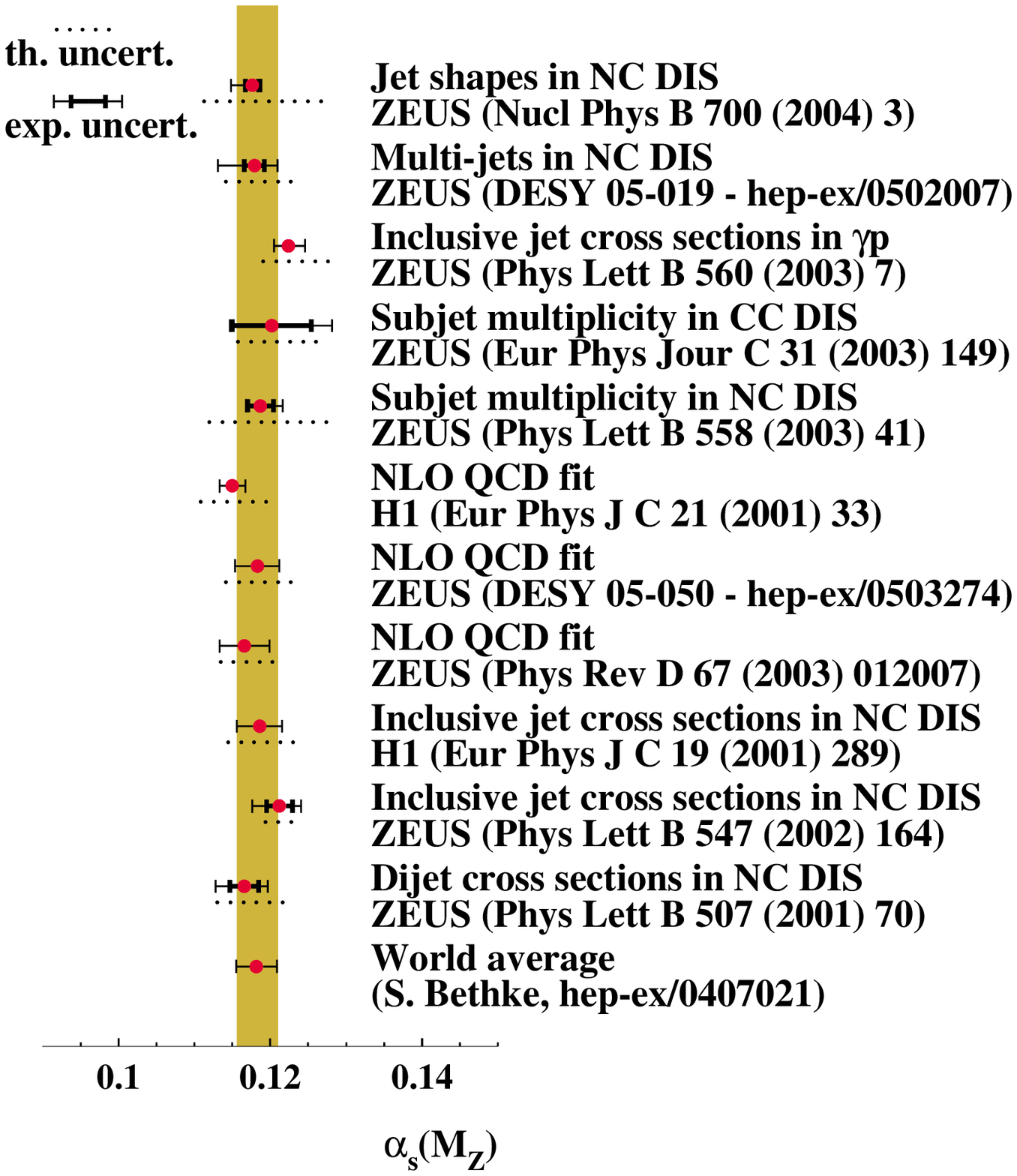,width=8.0cm}}
\put (5.0,-0.5){\epsfig{figure=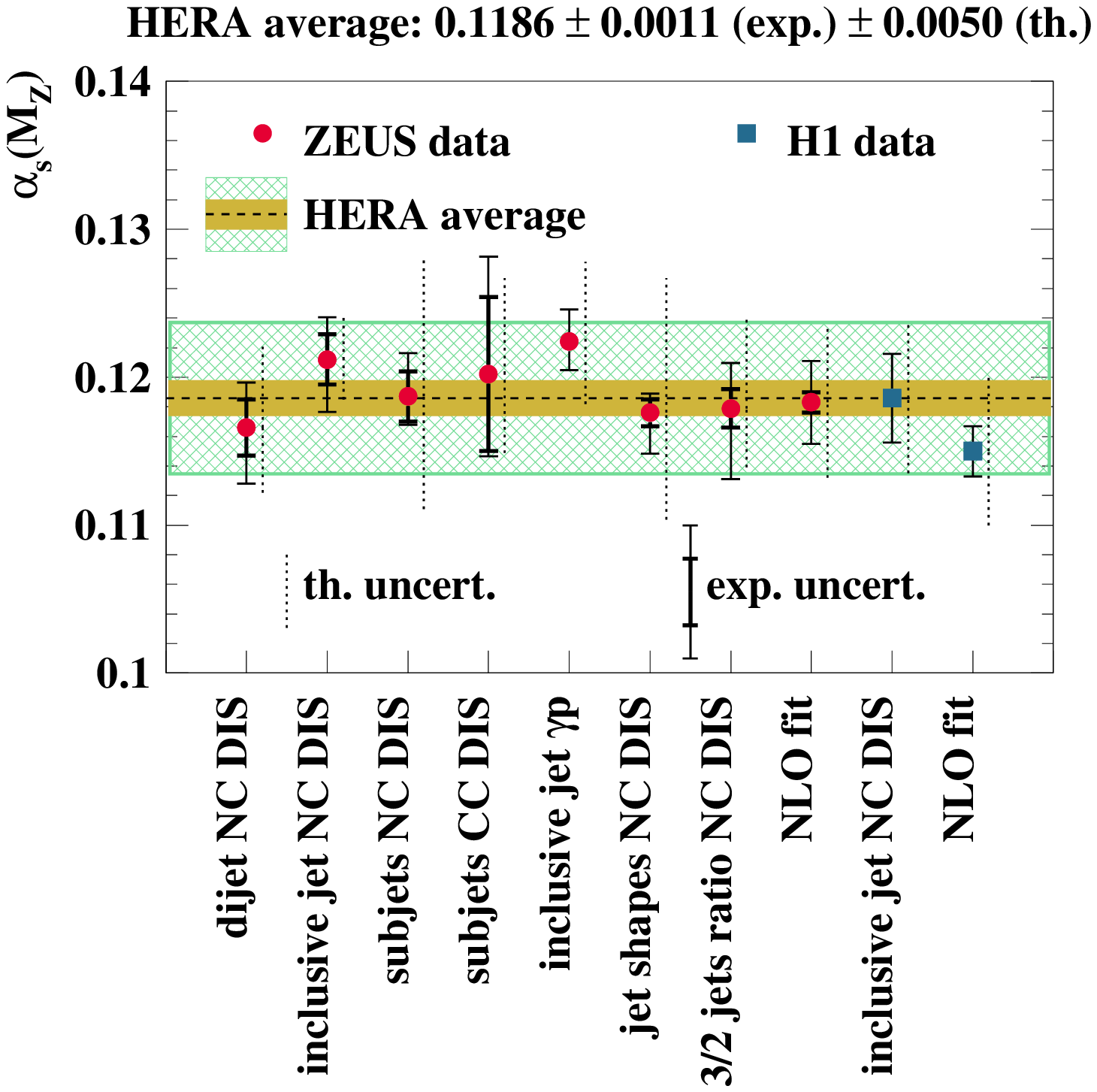,width=8.0cm}}
\put (-2.5,6.0){\bf\small (a)}
\put (5.0,6.0){\bf\small (b)}
\end{picture}
\caption{(a) Summary of $\asz$ determinations at HERA compared with
the world average; (b) $\asz$ determinations at HERA compared with
the HERA average.
\label{fig1}}
\end{figure}

The strong coupling constant, $\as$, is one of the fundamental
parameters of QCD. However, its value is not predicted by the theory
and must be determined by experiment. Many precise and consistent
determinations of $\as$ from diverse phenomena underlie the success of
perturbative QCD (pQCD). At HERA, $\as$ has been determined from many
observables, which include jet cross sections and structure functions,
by the H1 and ZEUS Collaborations. All the available
determinations~\cite{dijetzeus,incdiszeus,incdish1,nlofitzeus,nlofitzeusn,nlofith1,subnczeus,subcczeus,incgpzeus,multijetzeus,shanczeus}
are shown in Fig.~\ref{fig1}a. They are in good agreement with each
other and are consistent with the current world average
($\overline\asz^{WA}=0.1182\pm 0.0027$~\cite{wa}). These
determinations, most of which come from observables which involve jet
algorithms, lead to determinations of $\as$ some of which are as
precise as those from more inclusive measurements. The uncertainty in
these determinations is dominated by the theoretical contributions,
which amount to $4\%$ for jet cross sections and fits of structure
functions and $8\%$ for the internal structure of jets, whereas the
experimental uncertainties amount to $\sim 3\%$.

\section{Averaging the $\as$ determinations from HERA}

To make a proper average of all these diverse measurements, the
correlation among the different determinations has to be taken into
account. The experimental contribution to the uncertainty due to that
of the energy scale of the jets, which is the dominant source in
the jet measurements, is correlated among the determinations
from each experiment. On the theoretical side, the uncertainty
coming from the proton parton distribution functions (PDFs) is
certainly correlated whereas that coming from the hadronisation
corrections is only partially correlated. The uncertainty coming from
the terms beyond NLO is correlated up to a certain, a priori unknown,
degree; since these uncertainties are dominant, special care must be
taken in the treatment of these uncertainties when making an average
of the HERA determinations.

Several methods have been used to obtain an average value of $\as$ from
the HERA measurements and its uncertainty. Using a naive method in
which all uncertainties are assumed to be uncorrelated, the average
value and its uncertainty are:

\vspace{0.1cm}
\centerline{$\overline\asz=0.1188\pm 0.0020\ ({\rm ZEUS+H1}).$}
\vspace{0.1cm}

The second method used is that developed by
M. Schmelling~\cite{schmelling} to average correlated data when
correlations are present but hard to quantify. 
In this method, an error-weighted average and an
optimised correlation error were obtained from the error covariance
matrix by assuming an overall correlation factor between the total
errors of all measurements; the overall factor was determined by the
condition that the overall $\chi^2/{\rm dof}$ is equal to unity.
First, an error-weighted average was done separately for
the ZEUS and H1 measurements, and then the two averages were
combined:

\vspace{0.1cm}
\centerline{$\overline\asz=0.1196\pm 0.0060\ ({\rm ZEUS})\hspace{1cm}{\rm and}\hspace{1cm}\overline\asz=0.1166\pm 0.0053\ ({\rm H1}),$}
\vspace{0.1cm}

\centerline{$\overline\asz=0.1188\pm 0.0057\ ({\rm ZEUS+H1}).$}
\vspace{0.1cm}

The averages from the ZEUS and H1 determinations are compatible within
the uncertainties. The uncertainty of the combined average is 
$\sim 5\%$. 
This procedure gives rise to relatively large uncertainties when there
are large correlations among some of the measurements, as it is the
case here. To overcome this effect, the method has been repeated by
restricting to the most accurate measurements~\cite{bethke}. The
result of applying the procedure to those measurements with a total
error
$\Delta\asz<0.006$~\cite{incdiszeus,incdish1,nlofitzeusn,nlofith1,incgpzeus}
is:

\vspace{0.1cm}
\centerline{$\overline\asz=0.1192\pm 0.0047\ (\Delta\as^i<0.006)\ ({\rm ZEUS+H1}).$}
\vspace{0.1cm}

Finally, a more reliable, but conservative, approach has been used in
which the known correlations from the determinations of $\as$ coming
from the same experiment were taken into account (``correlation
method''). The theoretical uncertainties arising from the terms beyond
NLO were assumed to be (conservatively) fully correlated.
Error-weighted averages were obtained separately for the ZEUS and H1
measurements:

\vspace{0.1cm}
\centerline{$\overline\asz=0.1200\pm 0.0023\ ({\rm exp.})_{-0.0049}^{+0.0058}\ ({\rm th.})\ ({\rm ZEUS}),$}
\vspace{0.1cm}

\centerline{$\overline\asz=0.1160\pm 0.0016\ ({\rm exp.})_{-0.0049}^{+0.0048}\ ({\rm th.})\ ({\rm H1}).$}
\vspace{0.1cm}

A HERA average was obtained by using the error-weighted average method
on the ZEUS and H1 averages, assuming the experimental uncertainties
to be uncorrelated and taking the overall theoretical uncertainty as the
linear average of its contribution in each experiment. As a result, the
average of the HERA measurements and its uncertainty are:

\vspace{0.1cm}
\centerline{$\overline\asz=0.1186\pm 0.0011\ ({\rm exp.})\pm 0.0050\ ({\rm th.})\ ({\rm ZEUS+H1}).$}
\vspace{0.1cm}

This average, together with the individual values considered, is shown
in Fig.~\ref{fig1}b. It is found to be in good agreement with the
current world average, which does not include any of these
determinations. The results of applying Schmelling's and the
correlation methods are very similar, giving confidence on the average
obtained and its estimated uncertainty.

\section{Energy-scale dependence of $\as$}

\begin{figure}[h]
\setlength{\unitlength}{1.0cm}
\begin{picture} (10.0,6.0)
\put (-3.0,-1.0){\epsfig{figure=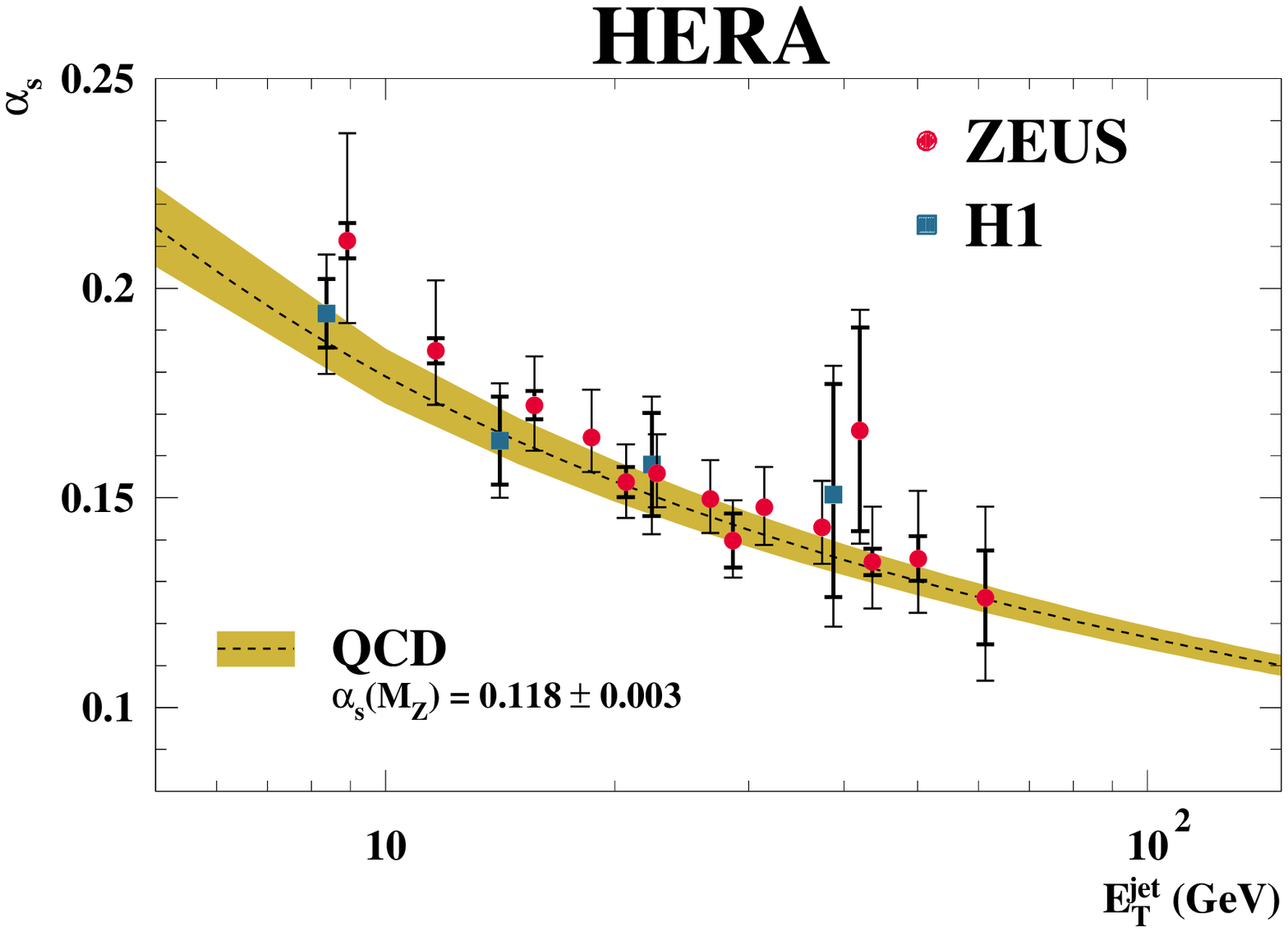,width=8.0cm}}
\put (5.0,-1.0){\epsfig{figure=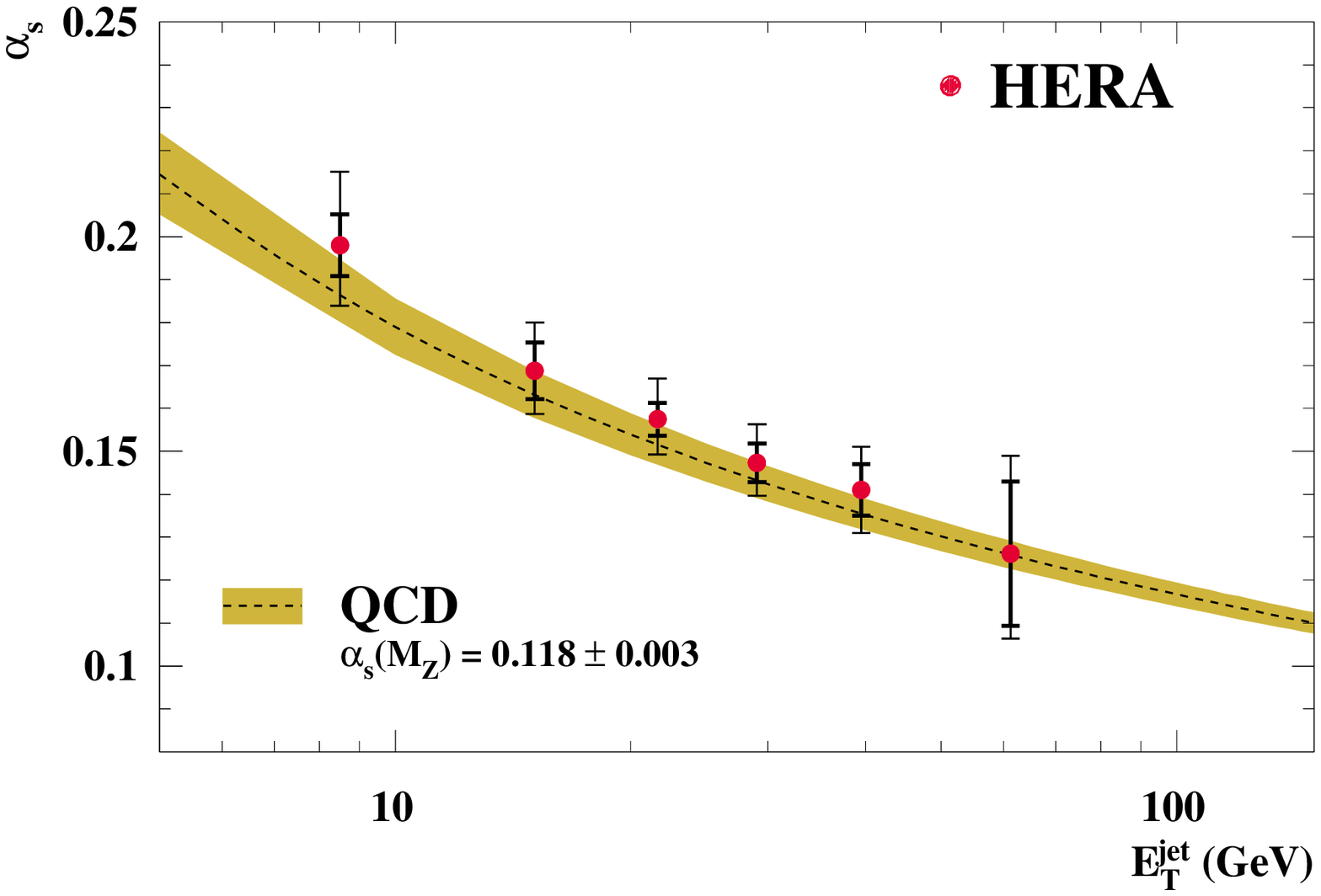,width=8.0cm}}
\put (3.6,4.6){\bf\small (a)}
\put (11.6,4.6){\bf\small (b)}
\end{picture}
\caption{(a) $\as$ as a function of $\etjet$ from H1 and ZEUS; (b)
  combined $\as$ as a function of $\etjet$ from HERA jet data. In both
  figures, the QCD prediction for the running of $\as$ is also
  shown. In (a), the inner error bars display the statistical
  uncertainties and the outer error bars display the systematic and
  theoretical uncertainties added in quadrature. In (b), the inner
  (outer) error bars show the experimental (theoretical) uncertainties.
\label{fig2}}
\end{figure}

The H1 and ZEUS Collaborations have tested the pQCD prediction for the
energy-scale dependence of the strong coupling constant by determining
$\as$ from the measured differential jet cross sections at different
$\etjet$~\cite{dijetzeus,incdiszeus,incdish1,incgpzeus}.
Figure~\ref{fig2}a shows the determinations of the energy-scale
dependence of $\as$ as a function of $\etjet$ from H1 and ZEUS. The
determinations are consistent with the running of $\as$ as predicted
by pQCD over a large range in $\etjet$.

The determinations of $\as(\etjet)$ from H1 and ZEUS at similar
$\etjet$ have been combined using the correlation method explained
above. The combined HERA determinations of the energy-scale dependence
of $\as$ are shown in Fig.~\ref{fig2}b, in which the running of $\as$
from HERA jet data alone is clearly observed.

\section{Summary}

A comprehensive average of $\as$ and its energy-scale dependence from
HERA data has been performed taking into account the known
correlations in each experiment and assuming conservatively that the
theoretical uncertainties arising from the terms beyond NLO are fully
correlated. The HERA average is

\vspace{0.1cm}
\centerline{$\overline\asz=0.1186\pm 0.0011\ ({\rm exp.})\pm 0.0050\ ({\rm th.}).$}
\vspace{0.1cm}
 
The experimental uncertainty of this average is $\sim 0.9\%$ and the
theoretical uncertainty amounts to $\sim 4\%$.
There is still room for improvement when the next-to-NLO (NNLO)
calculations needed for the determination of the PDFs are included and
when the NNLO calculations needed for jet-based observables are
finished.

\begin{theacknowledgments}
This work has been carried out in collaboration with Juan Terr\'on.
I would like to thank my colleagues from H1 and ZEUS for their help in
the preparation of this report. 
\end{theacknowledgments}

\end{document}